\documentclass[twocolumn]{aa}

\usepackage[dvipsnames]{xcolor}

\usepackage{hyperref}
\hypersetup{colorlinks=true,linkcolor=magenta,citecolor=blue,urlcolor=ForestGreen}

\usepackage{subcaption}

\graphicspath{{./}{figures/}}

\newcommand\editone[1]{\textcolor{black}{#1}}

\begin{document}


\title{Assessing the impact of background subtraction\\on a dust-free corona}

\author{Erika~Palmerio\inst{1}\corrauth{epalmerio@predsci.com} 
    \and Cooper~Downs\inst{1}\email{cdowns@predsci.com}
    \and Phillip~Hess\inst{2}\email{phillip.n.hess2.civ@us.navy.mil}
    \and Benjamin~J.~Lynch\inst{3}\email{bjlynch@ucla.edu}
    \and Ryder~Davidson\inst{1}\email{rdavidson@predsci.com}
    \and Ronald~M.~Caplan\inst{1}\email{caplanr@predsci.com}
    \and Jon~A.~Linker\inst{1}\email{linkerj@predsci.com}
    }

\institute{Predictive Science Inc., San Diego, CA 92121, USA
   \and U.S.\ Naval Research Laboratory, Washington, DC 20375, USA
   \and Department of Earth, Planetary, and Space Sciences, University of California, Los Angeles, CA 90095, USA
   }

\date{Received 13 August 2026 / Accepted 9 October 2026}

 
\abstract
{White-light observations of the solar corona contain contributions from both the electron-scattered K-corona and the dust-scattered F-corona. Because the diffuse F-corona dominates the measured intensity at altitudes of approximately $4\,R_{\odot}$ and above, background subtraction techniques are routinely employed to isolate the electron corona in coronagraph observations.}
{Since no direct observational ``ground truth'' for the K-corona exists, the amount of coronal signal that is inadvertently removed by background subtraction methods remains unknown. We aim to quantify, for the first time, the impact of background subtraction on the inferred large-scale structure and brightness of the solar corona.}
{We use a 32-day simulation based on a novel near-real-time, data-assimilative, time-evolving magnetohydrodynamic model to generate synthetic white-light observables. Because these synthetic observations are free from dust and instrumental stray-light contributions, they provide direct access to the ``true'' K-corona and enable a quantitative assessment of the effects of background subtraction.} 
{We find that background subtraction generally preserves the large-scale morphology of the electron corona, including the locations and shapes of major coronal structures. However, the procedure systematically modifies the relative brightness of different features and removes a non-negligible fraction of the K-coronal signal.}
{Although standard background subtraction techniques successfully recover the overall appearance of the electron corona, they can alter its quantitative photometric properties. These findings should be considered when interpreting coronal brightness measurements and deriving physical properties from processed white-light observations.}


\maketitle
\nolinenumbers


\section{Introduction} \label{sec:intro}

The large-scale structure of the tenuous solar corona has been fascinating humanity since ancient times. Before the development of technology and the advent of telescopes, the only occasions to observe the elusive solar atmosphere were represented by total solar eclipses, during which Luna occults the much brighter disc of the Sun, thus revealing structures such as streamers and prominences. Whether pushed by religious beliefs or by scientific curiosity, astronomers have been predicting \citep[e.g.,][]{steele2000, freeth2014}, drawing \citep[e.g.,][]{eddy1974, hayakawa2021}, and/or photographing \citep[e.g.,][]{hingley2001, pearson2011} the occurrence and appearance of eclipses throughout the centuries of human civilisation. However, total solar eclipses are rather rare events that take place approximately every 18 months somewhere on Earth's surface. With advancing technological progress and the advent of the Space Age, it has been possible to observe the corona more or less continuously via coronagraphs \citep[first invented in 1930; e.g.,][]{lyot1939}, which produce ``artificial eclipses'' by occulting the bright solar disc, thus allowing imaging of the photospheric light that is scattered in the corona. The first successful broadband white-light photograph of the solar corona outside of a total eclipse employing an external occulter took place from a rocket flight in 1963 and was reported by \citet{tousey1965}. This was followed by a series of coronagraphs flown on satellites to provide more regular observations of the corona, the timeline of which is well documented in \citet{howard2023}.

Today, several telescopes regularly observe the solar corona both from ground and from space, allowing monitoring of its large-scale structure and its transient phenomena. The Large Angle and Spectrometric Coronagraph \citep[LASCO;][]{brueckner1995} onboard the Solar and Heliospheric Observatory \citep[SOHO;][]{domingo1995} began providing nearly continuous synoptic coverage of the corona from the Sun--Earth L1 point in 1996. This was followed by similar instruments flown in 2006 on the twin Solar Terrestrial Relations Observatory \citep[STEREO;][]{kaiser2008} spacecraft as part of the Sun Earth Coronal and Heliospheric Investigation \citep[SECCHI;][]{howard2008}, orbiting the Sun close to 1~au. More recently, the Compact Coronagraph \citep[CCOR;][]{thernisien2025} series of instruments has brought white-light coronagraphy into the operational realm onboard the Geostationary Operational Environmental Satellites (GOES), in geostationary orbit, and the Space Weather Observations at L1 to Advance Readiness (SOLAR)-1 spacecraft, at the Sun--Earth L1 point. These were---and still are at the time of this writing---truly synoptic instruments, providing several images per hour every day and are the most relevant observational comparisons for this study.

The elusive solar corona, when observed in broadband white light, presents contributions from two main components of scattered photospheric light: the K-corona, due to electrons, and the F-corona, due to dust. While the K-corona corresponds to the ``true corona'' and displays structuring in the form of helmet streamers, pseudostreamers, coronal holes, and plumes, the diffuse F-corona dominates the measured intensity especially at altitudes ${\gtrsim}4\,R_{\odot}$ \citep[e.g.,][]{howard2019}. For this reason, revealing the appearance of the electron corona in coronagraph \editone{data} requires a multi-stage processing pipeline, including the production of \editone{background images} to remove the contribution of the dust corona. In fact, while the F-corona is much more intense than the K-corona, it is also more stable and visible changes tend to occur on much longer time scales than a single day \citep{stenborg2018, lamy2022}. Because K-corona structures are transient in nature, a stable signal in each day of coronagraph observations can be determined by taking a median intensity for each pixel over that given 24-hour interval, thus removing most of the noise fluctuations and the evolving component of the K-corona \citep[e.g.,][]{thompson2010}. To further reduce the impact of potential daily variations, the minimum of a series of these daily backgrounds (over a certain time period, usually 29 days) is employed as the so-called monthly minimum background. This background is subtracted off the ``raw'' coronagraph images taken within the time window to highlight the K-corona in a self-consistent manner. This approach is currently employed as the most standard background removal technique for both LASCO and SECCHI. \editone{The temporal window used to construct these backgrounds can also be adapted to the intended application. In particular, asymmetric windows, which rely more heavily on observations acquired before the image being processed, can be advantageous for real-time applications, where future data are not yet available \citep[e.g.,][]{thernisien2025}.}

\editone{Regardless of the temporal window adopted for background construction,} this method removes much, but not all, of the F-corona signal. Very close to the Sun, high-resolution imaging has revealed that the F-corona can be depleted by passing coronal mass ejections or the presence of high-speed streams related to coronal holes \citep{stenborg2023}. This means that, at least at some heights, the minimum values in the background will underestimate the typical F-corona. At the same time, it has been noted that background images may also have a residual K-coronal contribution to them (i.e., any part of the electron corona signal that persists over the considered time period, such as a quasi-stable helmet streamer), and that the significance of such artifacts may even depend on the phase of the solar cycle \citep{morrill2006}. To date, it has not been possible to quantify the amount of K-corona emission that remains during the background construction and how significant the resulting systematic uncertainties may be.

In this work, we take advantage of a 32-day-long, time-dependent magnetohydrodynamic (MHD) simulation to produce synthetic white-light images and to apply background removal techniques similar to those used for real coronagraph data. The advantage of such an approach is that, in the model, only the K-corona is present, hence we can effectively quantify how much structure is lost and/or altered by the generation and removal of background images. Our main goals in this effort include to determine the main features contributing to the final backgrounds, to evaluate the relative brightness of coronal structures before and after background removal, and to assess our results in the context of the ``ground-truth'' 3D configuration of the solar corona. \editone{This work is intended to provide a baseline for understanding the impact of background subtraction on the K-corona using MHD simulations, while laying the foundation for future studies that will incorporate increasingly realistic observing conditions, including the F-corona and instrumental effects.}


\section{Simulation overview and setup} \label{sec:setup}

The MHD simulation that we investigate in this work employs the Magnetohydrodynamic Algorithm outside a Sphere \citep[MAS;][]{mikic1999, mikic2018} code and was set up and run to predict the structure of the corona during the 2024 April 8 total solar eclipse\footnote{\url{https://www.predsci.com/eclipse2024}}. Detailed information on the model and its overall results is provided in \citet{downs2025}, hence we shall only provide a brief summary here. The novelty of this particular prediction effort is that it employs a near-real-time pipeline that acquires photospheric magnetic field maps, assimilates them into a flux transport model---specifically, the Open-source Flux Transport \citep[OFT;][]{caplan2025} code suite\footnote{\url{https://github.com/predsci/oft}}---and creates boundary conditions that evolve in time to drive a time-dependent MHD model of the solar corona up to 30\,$R_{\odot}$. This results in a coronal structure that emulates many well-known solar-like dynamics, including e.g.\ coronal rain, streamer disconnection events, small-scale eruptions, and coronal mass ejections. Hence, our simulation captures a number of transient features that illustrate both the small-scale dynamics and global-scale evolution of the solar corona over 32 days (2024 March 16 at 12:00~UT through April 17 at 19:00~UT), and is thus more than adequate for our purposes. Importantly, our analysis does not require the simulated corona to reproduce the precise timing, morphology, or evolution of individual observed features over this period. Rather, the objective is to generate physically realistic, time-dependent synthetic coronagraph imagery that captures the range of structures and dynamics expected in the real corona, providing an appropriate environment in which to investigate the effects of background subtraction.

From the simulation output, we perform line-of-sight integrations over the resulting 3D data cubes of coronal density (with hourly cadence) to derive plane-of-sky total and polarised brightness \citep[e.g.,][]{howard2009}, which we shall henceforth refer to as tB and pB, respectively. We choose to emulate the field of view of the SECCHI/COR2 telescope, i.e.\ covering the approximate range of 2.5--15\,$R_{\odot}$. This ensures that we capture the full range of coronal heights associated with interesting dynamics whilst keeping ``far enough'' from the outer boundary of the simulation's coronal domain. The viewpoint employed for the generation of white-light images corresponds to the real trajectory of the STEREO-A spacecraft over the duration of the simulation.

Throughout this work, we consider three synthetic coronagraph data sets and four background generation methods. The first and second sets of images correspond to time-dependent, white-light tB and pB observations throughout the simulation interval, respectively. The third data set is generated by considering the 3D density cube at the central time in the model (i.e., 2024 April 1 at 12:00~UT) and flying the STEREO-A trajectory through this ``static'' corona to recover synthetic tB observations. Regarding the backgrounds, we employ the ``traditional'' construction method that calculates the minimum of a series of 29 daily medians centred on the observation time (29d Med), and additionally use a simpler 29-day minimum (29d Min) as well as the same two methods for 15-day windows (15d Med and 15d Min, respectively). Since the generation of these backgrounds requires availability of data over 14 days before and after the time of interest and the simulation has a total duration of 32 days, we are able to produce background-subtracted images that span three days (i.e., 2024 March 31 through April 2).


\section{Results} \label{sec:results}

As mentioned in the previous section, we consider the mid point in time of the simulation (i.e., 2024 April 1 at 12:00~UT) and generate three daily backgrounds around this date for each of the four construction methods. In this section, we present the results of our analysis aimed at evaluating the amount of K-corona emission leaking into the backgrounds as well as the presence (or lack thereof) of artifacts manifesting themselves after background removal. We remark that, in all tB and pB images presented in this article, a simple radial filter (asymptotically approaching an $r^{3}$ dependence) has been applied to enhance emission at increasingly larger coronal heights.

\subsection{Overall appearance of the corona} \label{subsec:appearance}

The first step in our analysis consists of generating the various sets of background-subtracted images (described in Section~\ref{sec:setup}) and comparing them to the corresponding ``raw'' observables. The daily backgrounds produced and employed for each data set are presented and described further in Appendix~\ref{app:backgrounds}. Figure~\ref{fig:alltheimages} and its animated version show the three sets of raw imagery (time-dependent tB and pB as well as static tB) together with their corresponding background-subtracted images (minimum of daily medians, Med, and simple minimum, Min) for both the 15-day and 29-day series. We emphasise that the intensity scales shown in these panels are chosen solely for display purposes, as is commonly done in qualitative inspections of coronagraph data by researchers and forecasters. The raw tB images are all displayed using the same intensity range, while the background-subtracted tB images use a separate range because background removal substantially reduces the overall signal. Within the background-subtracted set, nevertheless, all tB images share the same intensity range to allow direct comparison between the different background subtraction methods. The same approach is used for pB, with intensity ranges adjusted to account for its lower (approximately two-thirds that of tB) overall brightness.

\begin{figure*}[p!]
\centering
\includegraphics[width=.7\linewidth]{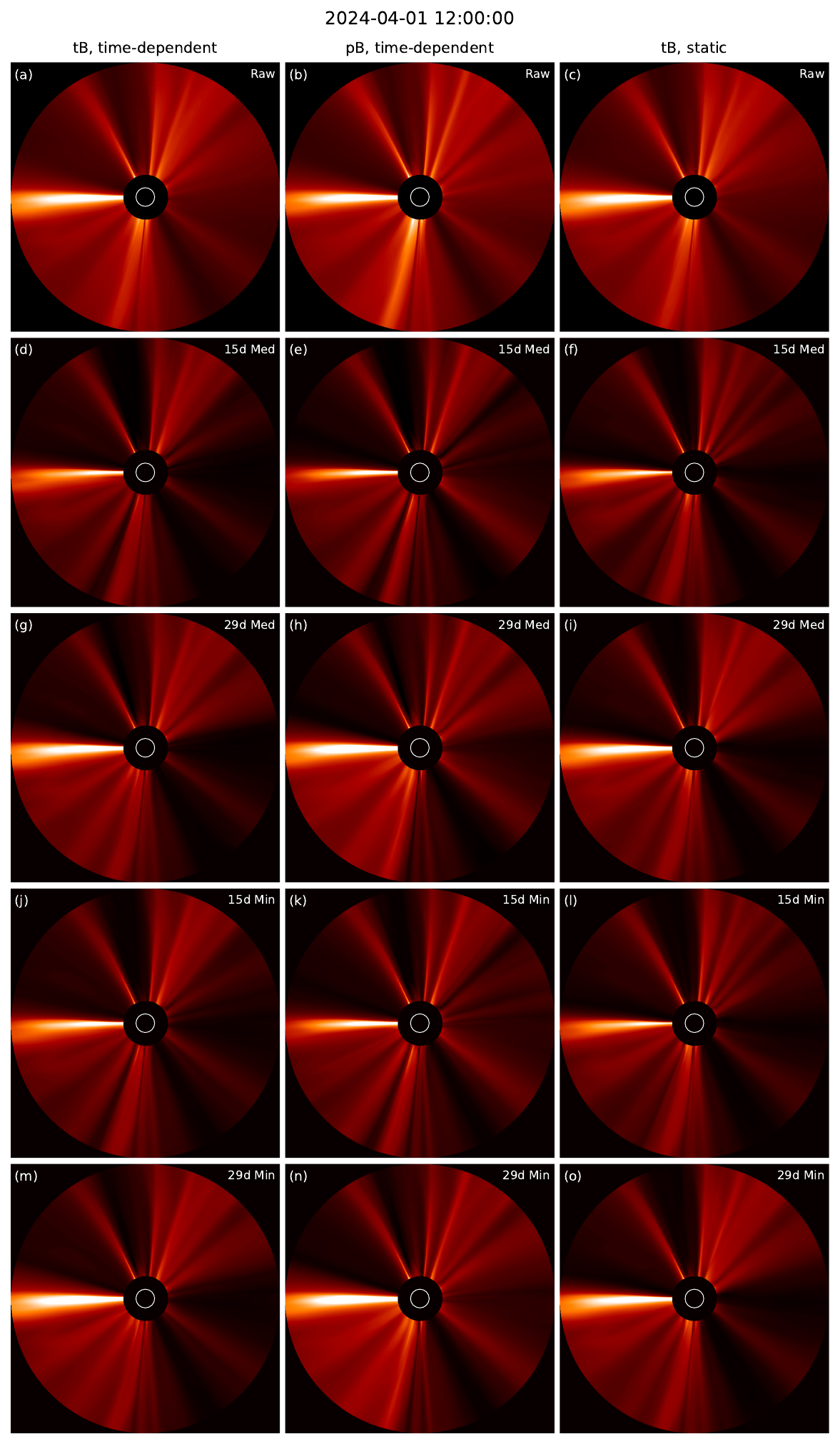}
\caption{Raw data and corresponding background-subtracted images for the 15-day and 29-day series (Med and Min) on April 1, 2024 at 12:00~UT. The left column displays time-dependent tB imagery, the middle column displays time-dependent pB imagery, and the right column displays static tB imagery. The raw tB and pB images are each displayed using independent intensity ranges, while all background-subtracted images of a given observable share a common intensity scale. The animated version of the figure shows the full three-day range (from March 31 to April 2, 2024) of hourly data considered in this work.}
\label{fig:alltheimages}
\end{figure*}

At first glance, it is evident that all the images in Figure~\ref{fig:alltheimages} depict the same instantaneous state of the solar atmosphere, indicating that the background subtraction procedure does not alter the structure of the corona in a manner that is immediately apparent to the eye. Nevertheless, a number of differences between the data sets become evident upon closer inspection: while streamers and other bright structures appear largely unchanged, the regions displaying fainter emission become noticeably darker. Furthermore, the 15-day backgrounds produce visibly larger deviations from the corresponding raw images than their 29-day counterparts. For example, all 15-day images exhibit a dark ``slice'' near position angle (PA\footnote{Note that the position angle \editone{(PA)} is measured counter-clockwise from solar north.}) $160^{\circ}$. This feature is substantially attenuated in the 29-day series and corresponds to a relatively bright region in the raw images.

Another striking difference is evident in the animated version of Figure~\ref{fig:alltheimages}. Because the sequence spans three consecutive days, it is possible to identify a clear ``jump’’ in the processed images whenever the background changes from one day to the next. This effect is considerably more pronounced for the 15-day series than for the 29-day series, indicating larger differences between consecutive daily backgrounds (see also Appendix~\ref{app:backgrounds}). This suggests that the traditional 29-day backgrounds, constructed over approximately one solar rotation, average out short-term coronal variability more effectively than the 15-day backgrounds. As a result, consecutive backgrounds evolve more smoothly in time, reducing artificial discontinuities in the background-subtracted images. \editone{We note that one possible way to mitigate these abrupt changes, whether large or small, is to interpolate between consecutive backgrounds to achieve smoother transitions across the images used for the subtraction. This approach is employed, for example, in the LASCO pipeline, where a 29-day, Med-like background is produced weekly and the intermediate days are obtained via linear interpolation \citep{morrill2006}. Similarly, the PUNCH data processing pipeline interpolates between daily F-corona models to generate a background for each data product \citep{deforest2026}. However, interpolation may also introduce additional artifacts, and its impact on the resulting K-coronal signal would need to be assessed. In the present study, given our highly idealised setup, we focus on intrinsic changes in the appearance of the corona---and, consequently, in the generated backgrounds---by independently calculating the backgrounds for each consecutive day without introducing additional processing steps.}

Overall, we do not identify substantial qualitative differences between the various data sets (time-dependent tB and pB as well as static tB). The two time-dependent data sets exhibit a remarkably similar behaviour throughout the interval considered, whilst the static tB images show somewhat larger deviations, as expected given that they do not include the temporal evolution of the corona. Over the remainder of this section, we shall quantify the similarities and differences identified through this visual inspection and assess how they depend on the choice of background subtraction method and input data.

\subsection{Loss in brightness} \label{subsec:brightness}

\begin{figure*}[p!]
\centering
\includegraphics[width=\linewidth]{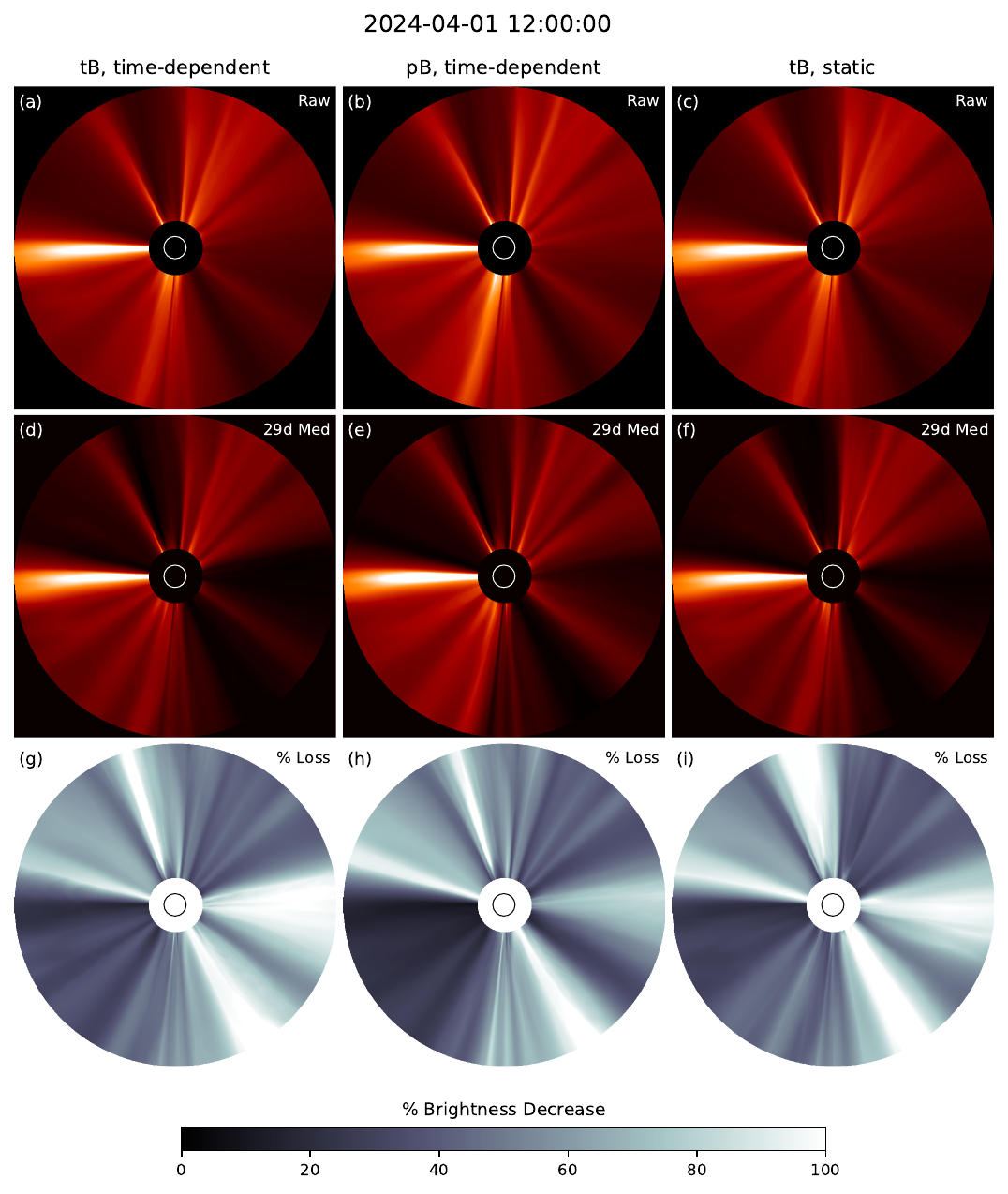}
\caption{Raw data and background-subtracted images for 29-day Med series, together with the corresponding pixel-by-pixel percentage loss in K-coronal brightness, on April 1, 2024 at 12:00~UT. The left column displays time-dependent tB imagery, the middle column displays time-dependent pB imagery, and the right column displays static tB imagery. The animated version of the figure shows the full three-day range (from March 31 to April 2, 2024) of hourly data considered in this work.}
\label{fig:percentmap_29dmed}
\end{figure*}

As discussed in Section~\ref{subsec:appearance}, visual inspection of the background-subtracted images suggests that regions devoid of streamers become considerably darker following background removal. To quantify this effect, we compute the pixel-by-pixel percentage loss in brightness relative to the true K-corona over the three-day interval considered in this work. Figure~\ref{fig:percentmap_29dmed} and its animated version show the resulting maps for the 29-day Med series for each of the three data sets. Although not shown here, the remaining three background series (29-day Min, 15-day Med, and 15-day Min) exhibit a qualitatively similar behaviour. Note that the radial vignette used on the raw and background-subtracted images to enhance coronal emission with increasing heliocentric distance has not been applied to the percentage-loss maps (although such a spatial correction would mathematically cancel out regardless).

The results displayed in Figure~\ref{fig:percentmap_29dmed} and its accompanying animation reveal that the brightness removed by the background subtraction procedure is far from spatially uniform. Instead, the fractional loss exhibits pronounced structuring with PA, closely following the large-scale morphology of the corona itself. As noted via qualitative inspection in Section~\ref{subsec:appearance}, bright streamer regions generally experience the smallest relative losses, whereas the darker areas between streamers, corresponding predominantly to open-field regions, undergo substantially larger fractional losses. Consequently, background subtraction preferentially suppresses emission in the intrinsically faint portions of the corona while leaving the brightest structures comparatively unaffected. This behaviour is consistent across all three input data sets, whose percentage-loss maps display nearly identical spatial distributions. 

Despite the similar spatial distributions, the magnitude of the brightness loss differs more or less significantly between the three input data sets. On average, the median percentage loss in the time-dependent pB images is approximately 90\% of that obtained from the corresponding time-dependent tB images, with values ranging between 85\% and 95\% over the three-day interval. Thus, background subtraction appears to consistently remove a smaller fraction of the K-coronal signal when applied to pB rather than to tB imagery. In contrast, the differences between the time-dependent and static tB data sets are more variable, and show no systematic offset. Over the temporal window of interest, the median percentage loss in the static tB images ranges from 93\% to 117\% of that in the corresponding time-dependent tB series, indicating that the use of a static corona can either slightly under- or overestimate the amount of K-coronal signal removed.

\subsection{Relative brightness of structures} \label{subsec:rings}

\begin{figure*}[t!]
\centering
\includegraphics[width=\linewidth]{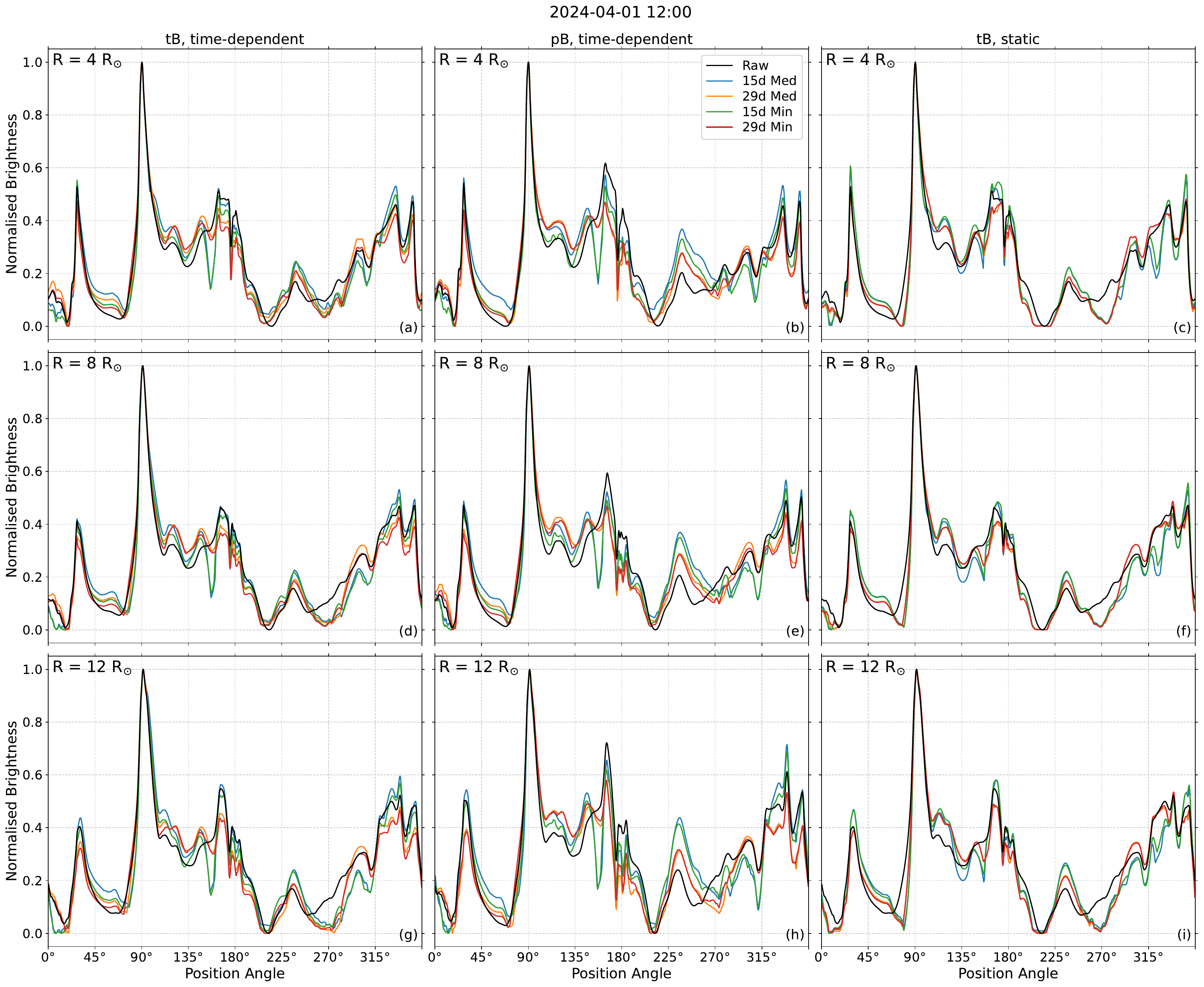}
\caption{Normalised brightness at three different coronal heights (4, 8, and 12\,$R_{\odot}$) with respect to position angle for the raw images shown against the corresponding background-subtracted ones. The different columns show the (left) time-dependent tB, (middle) time-dependent pB, and (right) static tB data sets. The animated version of the figure shows how emission at different coronal heights changes in time over the three days of hourly data considered in this work (from March 31 to April 2, 2024).}
\label{fig:ringplot_images}
\end{figure*}

While the analysis in Section~\ref{subsec:brightness} quantified the absolute loss of coronal brightness introduced by background subtraction, it does not indicate whether the relative brightness of different structures is preserved. This is an important aspect to consider, since the relative contrast between streamers, pseudostreamers, and coronal holes may affect interpretation of white-light observations. To investigate this, we compare the variation in coronal brightness with PA at fixed heliocentric distances. Specifically, for each raw and background-subtracted image, we extract annular profiles (``rings'') at heliocentric distances of 4, 8, and 12\,$R_{\odot}$. To facilitate comparison between the different data sets, we normalise each profile to its minimum and maximum values, thus emphasising changes in the relative brightness distribution rather than the absolute intensity. The resulting normalised PA profiles are shown in Figure~\ref{fig:ringplot_images}, and equivalent profiles extracted from the background images themselves are presented in Appendix~\ref{app:rings_bkg}.

Figure~\ref{fig:ringplot_images} and its accompanying animation first reveal that the changes in the relative brightness of coronal structures introduced by background subtraction are largely independent of heliocentric distance. In other words, when a given feature appears relatively brighter or fainter than another after background subtraction, this behaviour is preserved from 4 to 12\,$R_{\odot}$. The background-subtracted profiles therefore maintain essentially the same relative deviations from the corresponding raw profiles throughout the range of coronal heights considered. At the same time, the normalised profiles demonstrate that background subtraction largely preserves the overall brightness distribution around the Sun. The locations of the principal maxima and minima remain virtually unchanged, indicating that the large-scale morphology of the corona is retained. The primary effect of the subtraction is instead to modify the relative amplitudes of individual structures: while some bright peaks (corresponding to streamers) become more pronounced with respect to the raw profiles, others are systematically reduced, with analogous changes occurring in the darker regions between them (corresponding to coronal holes). This indicates that the background subtraction procedure does not simply rescale the coronal brightness uniformly, but redistributes the relative contrast between different coronal features.

The figure also shows that, for a given input data set, the 29-day Med and Min profiles are generally much more similar to one another than to the corresponding 15-day profiles, and likewise for the two 15-day methods, as is also evident from the respective background profiles discussed in Appendix~\ref{app:rings_bkg}. This suggests that the duration of the background window has a considerably larger influence on the relative brightness distribution than the choice of background construction method, at least for the methods and durations considered here. Finally, the animated version of Figure~\ref{fig:ringplot_images} shows that the two time-dependent data sets (tB and pB) maintain a similar correspondence with their respective raw profiles throughout the three-day interval. In contrast, the apparent deviations of the static tB profiles become progressively larger away from 2024 April 1 at 12:00~UT, the single snapshot from which the static corona is constructed. This behaviour is expected, since the static imagery cannot reproduce the temporal evolution of the large-scale coronal structure captured by the time-dependent simulation. Nevertheless, when the background-subtracted static tB images are compared with their corresponding static raw profiles, they exhibit qualitatively similar modifications to the relative brightness distribution as the time-dependent data sets. Thus, the increasing disagreement with the time-dependent cases primarily reflects the evolution of the reference corona itself, rather than a degradation of the background subtraction procedure.


\section{Discussion} \label{sec:discussion}

In the previous section, we quantified the impact of the four background-construction methods (15-day and 29-day Med and Min) on the inferred structure of the K-corona for the three data sets considered (time-dependent tB, time-dependent pB, and static tB). In this section, we investigate the physical origin of the observed differences. Specifically, we examine how observations acquired at different times contribute to the daily backgrounds, quantify the loss of K-coronal signal and the preservation of coronal structure, and interpret these results in the context of the 3D structure and temporal evolution of the solar corona.

\subsection{Contributions to the backgrounds} \label{subsec:minmaps}

\begin{figure*}[p!]
\centering
\includegraphics[width=.8\linewidth]{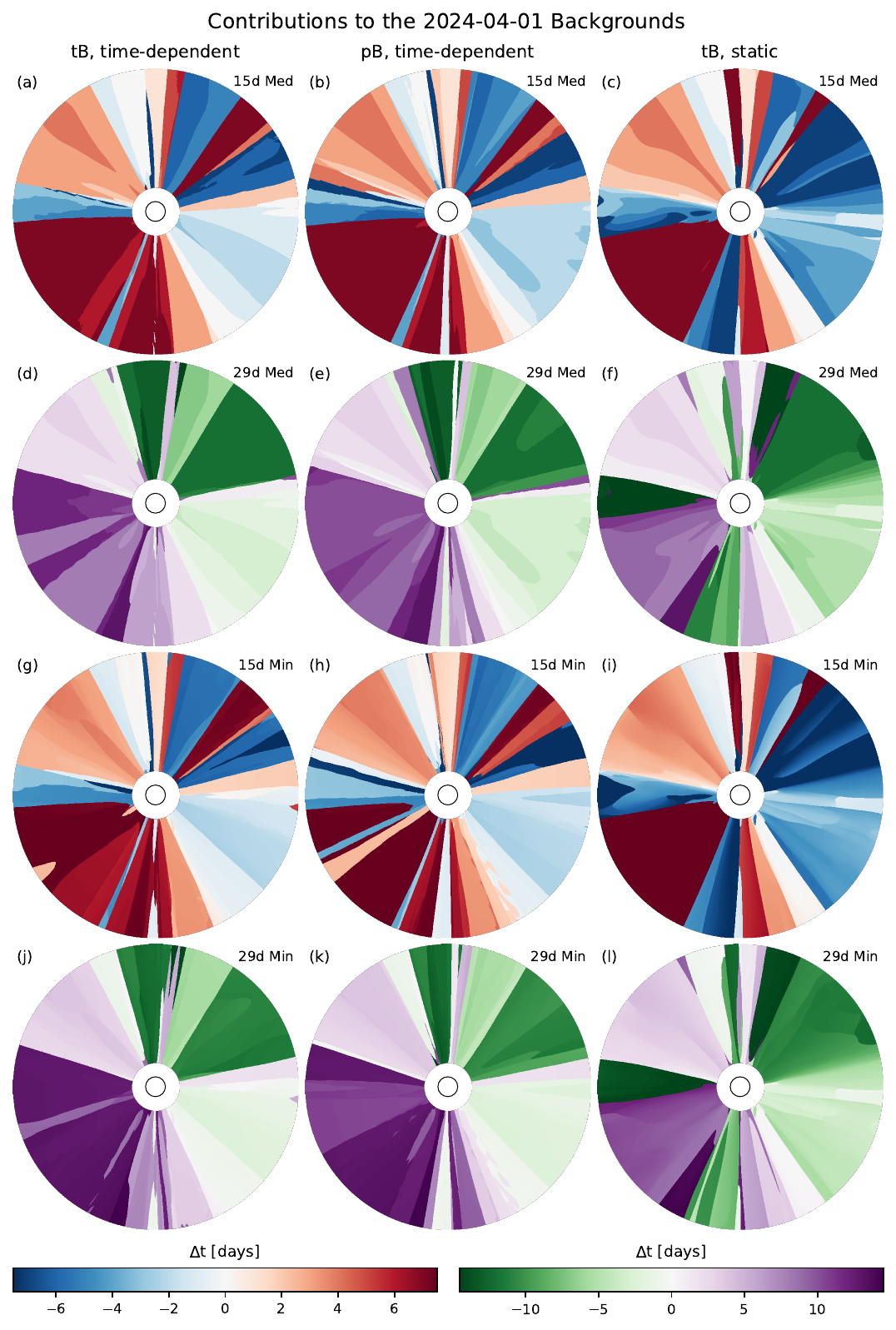}
\caption{Maps showing the time from which the minimum brightness value for each pixel is selected during the construction of the daily background. Rows display the four background construction methods (15-day Med, 29-day Med, 15-day Min, and 29-day Min), whilst columns show backgrounds derived from (left) time-dependent tB, (middle) time-dependent pB, and (right) static tB imagery. The colours indicate the temporal offset, $\Delta t$, in days relative to the background date (2024 April~1), with negative (positive) values corresponding to observations acquired before (after) this date. The animated version of the figure shows the equivalent maps for the full three-day interval considered in this work.}
\label{fig:minmaps_20240401}
\end{figure*}

Since background maps are constructed from observations spanning several days, the emission assigned to a given pixel depends on both the background construction method and the temporal evolution of the corona (arising from both its intrinsic dynamics and its apparent evolution due to solar rotation) over the chosen interval. For any background construction method, the value assigned to a given pixel is determined by the temporal distribution of brightness values within the selected background window. Physically, emission minima are expected to originate preferentially from times when the corresponding line of sight intersects relatively faint coronal regions, such as open-field areas associated with coronal holes. To investigate how the choice of background window and construction method influences the resulting backgrounds, we examine ``minimum maps'', in which each pixel is coloured according to the time from which its minimum brightness value is selected ($\Delta t$, in days relative to the background date). Figure~\ref{fig:minmaps_20240401} presents these maps for the backgrounds on 2024 April~1, whilst the accompanying animation shows their evolution over the full three-day interval considered in this work. Note that, by construction, the Med maps contain only integer values, as each pixel is associated with the median of the daily minima and therefore maps to a single day. In contrast, the Min maps may assume fractional-day values, since each pixel is selected directly from an individual coronagraph image with a precise timestamp.

Figure~\ref{fig:minmaps_20240401} immediately reveals a clear difference between the 15-day and 29-day backgrounds for the two time-dependent data sets. The 29-day maps are dominated by two broad regions of predominantly positive and negative temporal offsets, whereas the corresponding 15-day maps are divided into a larger number of narrower sectors in which positive and negative values alternate more frequently with PA. In contrast, both the 15-day and 29-day maps derived from the static tB data set display a similar alternating sector pattern, resembling that of the time-dependent 15-day backgrounds. Moreover, the two averaging windows considered are themselves in much closer agreement, with the sign of $\Delta t$ remaining the same over most PAs in both the Med and Min cases. A second noteworthy result is the close correspondence between the Med and Min maps. Although small differences in the magnitude of $\Delta t$ are apparent, the spatial distribution and sign of the temporal offsets are nearly identical for a given background window and input data set. This indicates that, despite their different implementations, the two methods are effectively equivalent in terms of the regions of the temporal window from which they draw their contributions.

The temporal evolution of the regions contributing the minimum emission can be examined in the animated version of Figure~\ref{fig:minmaps_20240401}. Consistent with the results presented in Section~\ref{sec:results}, the 15-day backgrounds evolve much more rapidly than their 29-day counterparts. In the time-dependent data sets, large regions of the 15-day maps frequently change the sign of $\Delta t$ from one day to the next, whereas the 29-day maps retain the same overall distribution of positive and negative temporal offsets, with the evolution consisting primarily of modest changes in the magnitude of $\Delta t$. In contrast, the 29-day static tB backgrounds exhibit a temporal behaviour more akin to that of the 15-day series. However, this is likely an expected consequence of the static approximation: because the underlying corona is fixed in time, the same large-scale structures reappear after approximately one solar rotation (aside from minor changes in spacecraft position). Consequently, rapid transitions between large positive and negative values of $\Delta t$ primarily reflect the recurrence of the same coronal configuration within the 29-day background window, rather than genuine temporal evolution of the corona.

A plausible explanation for the greater temporal stability of the 29-day backgrounds lies in the duration of the construction window itself. Over approximately one solar rotation, coronal structures at a given heliolatitude are observed crossing both the eastern and western limbs of the plane of sky. As a result, the faintest emission associated with a given structure has two opportunities to contribute to the background, making the selected minima less sensitive to the precise timing of the construction window or to short-term coronal evolution. In contrast, over a 15-day interval coronal structures may only traverse one limb, hence the selected minima have a stronger dependence on the start and end times of the background construction window, naturally producing the larger day-to-day variability that is observed.

\subsection{Fidelity of the inferred corona} \label{subsec:fidel}

\begin{figure*}[t!]
\centering
\includegraphics[width=.99\linewidth]{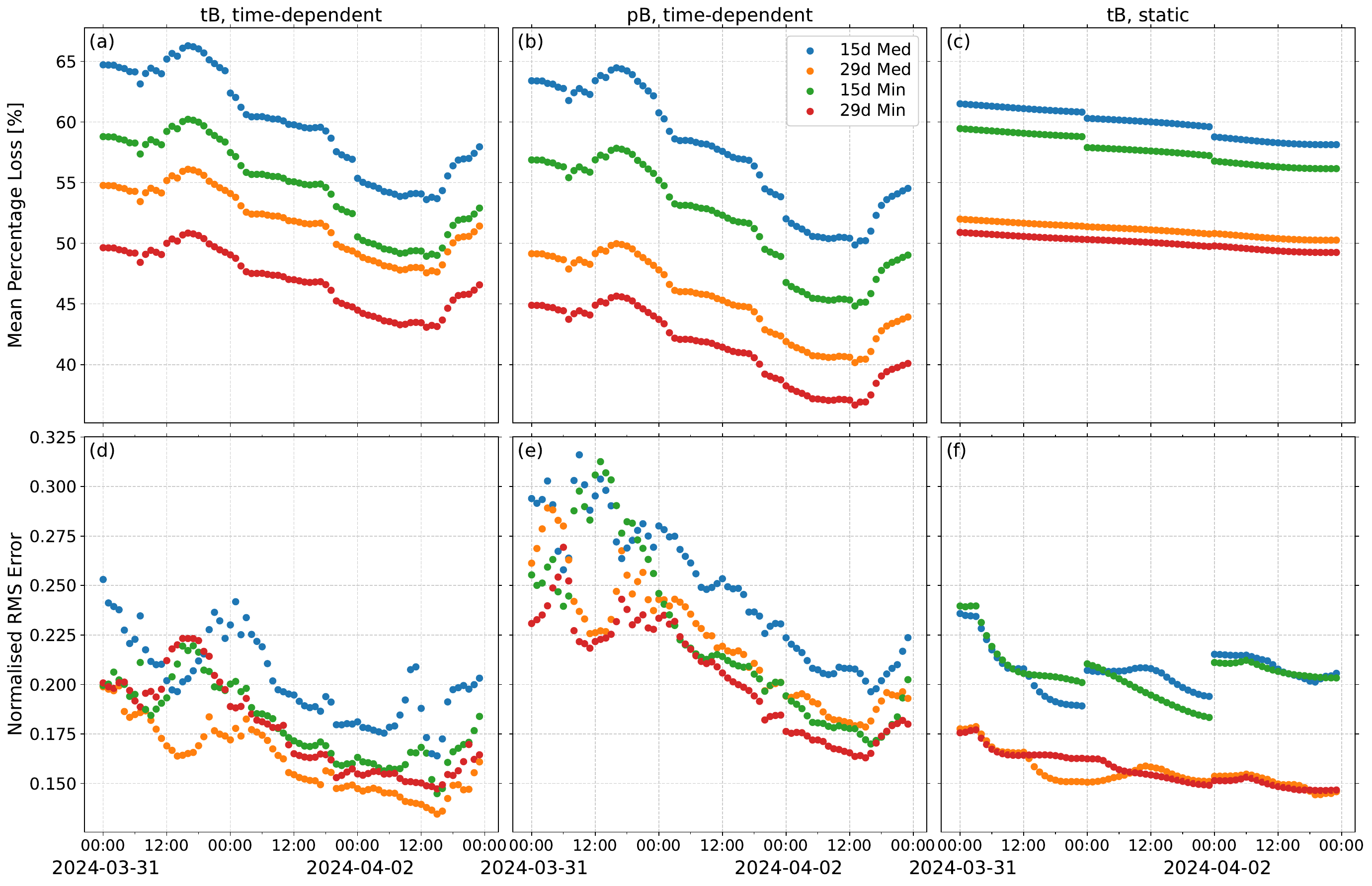}
\caption{Time series of the metrics applied to the annular brightness profiles at 8\,$R_{\odot}$ introduced in Section~\ref{subsec:rings}, shown for the (left) time-dependent tB, (middle) time-dependent pB, and (right) static tB data sets. Top: Mean percentage loss (MPL), quantifying the fraction of the total coronal signal removed by background subtraction. Bottom: Normalised root-mean-square error (NRMSE), quantifying the degree to which the relative brightness distribution of coronal structures is preserved after background subtraction.}
\label{fig:ring_timeseries}
\end{figure*}

Having examined the effects of background subtraction using 2D images and annular brightness profiles, we now summarise these results using two complementary metrics: the mean percentage loss (MPL), which quantifies the fraction of K-coronal signal removed:
\begin{equation}
    \mathrm{MPL} = 100 \cdot \left( 1 - \frac{\sum I_\mathrm{sub}}{\sum I_\mathrm{raw} } \right) \, ,
    \label{eq:mpl}
\end{equation}
and the normalised root-mean-square error (NRMSE), which quantifies how well the relative brightness distribution of coronal structures is preserved:
\begin{equation}
    \mathrm{NRMSE} = \sqrt{\frac{\sum (\tilde{I}_\mathrm{raw}-\tilde{I}_\mathrm{sub})^{2} }{\sum \tilde{I}^{2}_\mathrm{raw}}}\, .
    \label{eq:nrmse}
\end{equation}
Note that, in the equations above, $I_\mathrm{raw}$ and $I_\mathrm{sub}$ denote the intensities of the raw and background-subtracted images (in units of mean solar brightness), respectively, whilst $\tilde{I}_\mathrm{raw}$ and $\tilde{I}_\mathrm{sub}$ indicate their corresponding normalised quantities (with unitless values in the range [0, 1]). Since the previous sections showed that these effects vary primarily with PA rather than heliocentric distance, we restrict this analysis to the annular profiles at 8\,$R_{\odot}$ introduced in Section~\ref{subsec:rings}. Figure~\ref{fig:ring_timeseries} shows the temporal evolution of both metrics over the three-day interval for each background-construction method and input data set.

Regarding the MPL metric (top row of Figure~\ref{fig:ring_timeseries}), all three data sets (time-dependent tB and pB as well as static tB) exhibit the same ordering from lowest to highest percentage signal loss: 29-day Min, 29-day Med, 15-day Min, and 15-day Med. As noted in Appendix~\ref{app:backgrounds}, the minimum of a series of daily medians is necessarily greater than or equal to the direct minimum over the same interval, hence the lower MPL values obtained with the Min method are expected. Likewise, extending the background window from 15 to 29 days can only decrease (or leave unchanged) the minimum brightness selected at each PA, resulting in systematically lower MPL values for the 29-day background-subtracted images. As already discussed in Section~\ref{subsec:brightness}, the time-dependent pB data consistently exhibit lower percentage losses than the corresponding tB curves. This may be due to the fact that the pB signal is more localised around the Thomson surface, whilst tB receives contributions from a much broader range of distances \citep[e.g.,][]{deforest2013}. Consequently, coronal structures observed in pB contribute over a narrower range of viewing geometries and times, reducing the amount of K-coronal emission incorporated into the backgrounds and therefore the fraction of signal removed during subtraction. Interestingly, the separation between the four background construction methods is considerably larger for the two time-dependent data sets than for the static tB case. In the latter, not only are the differences between the 15-day and 29-day windows reduced, but the Med and Min methods also produce much more similar MPL values. This behaviour is also expected because, in the absence of intrinsic coronal evolution (including transients), the distribution of brightness values is expected to change less in time. Consequently, extending the background interval or replacing the median of the daily minima with the direct minimum has a comparatively smaller influence on the resulting background.

In contrast, the NRMSE metric (bottom row of Figure~\ref{fig:ring_timeseries}) exhibits no specific ordering of the four background construction methods. The relative errors vary with time, and the corresponding curves intersect repeatedly throughout the three-day interval. Nevertheless, the 29-day backgrounds generally produce lower NRMSE values than their 15-day counterparts, a trend that is particularly evident for the static tB data set. A second noteworthy result is that, unlike the MPL metric, the relative performance of the Med and Min methods is not consistent across the different data sets and background windows. For the static tB series, the two methods produce broadly comparable NRMSE values throughout the interval. In the time-dependent data sets, however, the best-performing method depends on both the input observable and the duration of the background window. For the 15-day backgrounds, the Min method generally yields lower NRMSE values than Med for both tB and pB. For the 29-day backgrounds, instead, Med performs slightly better for tB, whereas Min continues to produce the lower errors for pB. Finally, the time-dependent tB series exhibits substantially lower NRMSE values than the corresponding pB set throughout the interval considered. Thus, although the pB backgrounds retain a larger fraction of the absolute K-coronal signal (as reflected by the MPL metric), the relative brightness distribution of coronal structures is reproduced more faithfully in tB. This behaviour likely reflects again the different Thomson-scattering geometries of the two observables mentioned above. Due to the broader contribution to the tB signal, the resulting line-of-sight averaging produces a smoother brightness distribution that is less sensitive to the small changes introduced by background subtraction, despite a larger fraction of the total K-coronal signal being removed.

\subsection{Comparison with the 3D structure of the corona} \label{subsec:3d}

\begin{figure*}[t!]
\centering
\includegraphics[width=\linewidth]{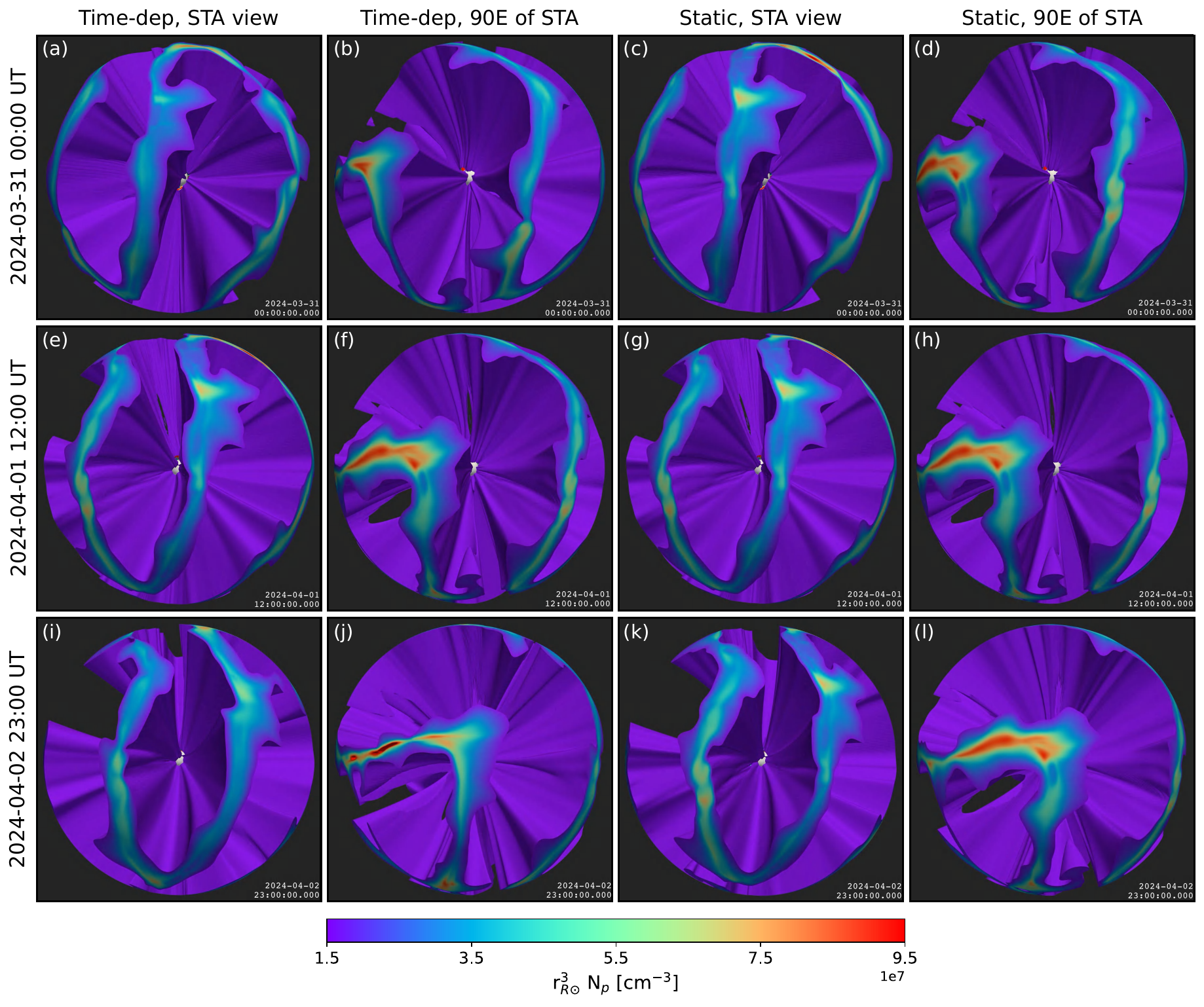}
\caption{3D visualisations of the modelled coronal density between 1.5 and 15\,$R_{\odot}$, illustrating the location of the streamer belt around the Sun. The columns present, from left to right, the time-dependent corona as viewed from STEREO-A and from a vantage point $90^{\circ}$ east of STEREO-A, followed by the corresponding static representations from the same two viewpoints. The rows correspond to the beginning, middle, and end of the three-day interval analysed in this work. The animated version of the figure extends the same four viewpoints over the full 32-day simulation.}
\label{fig:3dcorona}
\end{figure*}

One final aspect to investigate is how the performance of the background subtraction methods relates to the underlying 3D structure of the corona. Figure~\ref{fig:3dcorona} presents volume renderings of the modelled coronal density between 1.5 and 15\,$R_{\odot}$ for both the time-dependent and static data sets, viewed from the STEREO-A perspective and from a vantage point $90^{\circ}$ east of STEREO-A. \editone{For each rendering, the corresponding MAS density data cube is restricted to the radial range of 1.5--15\,$R_{\odot}$ and scaled by $r^3$ to compensate for the rapid radial decrease in coronal density. The volume is then rendered on a rectilinear grid using this scaled density, with values clipped to visually highlight the dense streamer structures associated with the heliospheric current sheet.} The \editone{resulting} visualisations reveal a highly tilted and warped streamer belt, characteristic of solar maximum conditions. Over the three-day interval analysed in this work, the large-scale morphology of the heliospheric current sheet remains broadly unchanged---although individual streamers exhibit modest evolution and small changes in shape and position, the overall topology of the corona appears generally well preserved. The accompanying animation, however, shows that over the full 32-day simulation, substantial differences develop between the static and time-dependent coronae as the streamer belt evolves in response to the continuously changing magnetic configuration of the photosphere. These changes include both the emergence and disappearance of streamer structures as well as noticeable distortions of the global current sheet geometry and morphology, illustrating that the intrinsic evolution of the corona becomes increasingly important over timescales approaching a solar rotation.

Nevertheless, many of the large-scale density structures that contribute to the constructed backgrounds remain common to both the time-dependent and static data sets over a substantial fraction of the 32-day simulation. In particular, the broad projected regions of comparatively low density, which preferentially contribute to the background minima identified in Figure~\ref{fig:minmaps_20240401} and its accompanying animation, are found at similar position angles in both cases. For example, the extended low-density region centred around PA $50^{\circ}$ in the STEREO-A view on 2024 April~2 at 23:00~UT (panels (i) and (k) of Figure~\ref{fig:3dcorona}) is present in both simulations and corresponds to one of the principal contributors to the backgrounds at those locations. Likewise, the animation accompanying Figure~\ref{fig:3dcorona} reveals that the projected low-density region near PA $300^{\circ}$ on March~21, together with the large cavity occupying much of the southeastern quadrant around April~12, are evident in both the time-dependent and static coronae despite the noticeable evolution of the surrounding streamer belt. These examples suggest that, although the detailed morphology of individual streamers evolves continuously, the large-scale projected distribution of dense and rarefied coronal plasma remains sufficiently similar over several days that the same broad structures continue to dominate the background construction (see also Figure~\ref{fig:backgrounds_20240401}). This may explain why the static backgrounds reproduce the qualitative behaviour of the time-dependent cases considerably better than might be expected from a non-evolving corona.

Indeed, it is worth remarking that the interval analysed here lies close to solar maximum, when the heliospheric current sheet is highly warped and the global coronal magnetic field evolves comparatively rapidly. One might therefore expect intrinsic coronal evolution to substantially degrade the performance of the static approximation. Instead, our results demonstrate that, over timescales of approximately a solar rotation, the large-scale projected density distribution remains sufficiently stable that the resulting ``static'' backgrounds are qualitatively similar to those obtained from the fully time-dependent simulation. Whether this conclusion also holds during solar minimum remains to be determined---the streamer belt is considerably less warped and evolves significantly more slowly during that phase of the solar cycle, suggesting that static approximations may perform even better, although the overall simpler heliospheric current sheet morphology may also alter the relative contributions of individual structures to the constructed backgrounds.


\section{Summary and conclusions} \label{sec:conclusions}

In this work, we have investigated the extent to which background subtraction applied to white-light coronagraph observations inadvertently subtracts K-coronal signal together with the intended F-corona contribution. Under the idealised assumption of a perfectly removed F-corona, any residual signal incorporated into the constructed backgrounds originates solely from the K-corona. We have exploited this scenario by analysing synthetic white-light imagery derived from the 32-day, near-real-time, data-assimilative MHD simulation of \citet{downs2025}, which contains only the K-coronal component. This allows, for the first time, a direct quantification of the effects of background subtraction against a known ground truth.

We have employed three input data sets (time-dependent tB, time-dependent pB, and static tB) together with four background construction methods (minimum of daily medians, Med, and direct minimum, Min, each applied over 15- and 29-day windows) to evaluate the impact of background subtraction over the three-day interval for which all backgrounds could be generated. We find that the overall morphology of the K-corona is remarkably well preserved following background subtraction. The largest coronal structures, including streamers, pseudostreamers, and open-field regions, remain readily identifiable, although quantitative analyses reveal systematic changes in both the absolute and relative brightness of different structures.

We found that constructing backgrounds over a 29-day window, as is commonly done for real coronagraph observations, consistently removes less total K-coronal signal and generally provides a more faithful recovery of the relative brightness distribution of coronal structures than the corresponding 15-day backgrounds. This demonstrates that sampling approximately one full solar rotation provides a more robust estimate of the background, despite the greater amount of intrinsic coronal evolution encompassed by the longer interval. Furthermore, we found that background subtraction applied to pB imagery removes systematically less K-coronal signal than for tB (by approximately 5\%), although the corresponding tB images generally preserve the relative brightness distribution of coronal structures slightly more faithfully.

We also noted that, although the Min method removes less total K-coronal signal than Med by construction, the two approaches perform similarly in preserving the relative brightness of streamers, pseudostreamers, and open-field regions. This is an encouraging result, as practical background subtraction applied to real coronagraph observations must rely on Med-like approaches to mitigate the influence of instrumental noise, cosmic ray hits, and other image artifacts, which would otherwise have a disproportionate impact on backgrounds constructed using the direct minimum. Since the synthetic observations analysed here are free from such effects, \editone{the relative performance of the Med and Min methods under more realistic observing conditions remains to be investigated.}

An additional noteworthy result is that backgrounds constructed from synthetic coronagraph images generated by ``flying'' a virtual spacecraft through a single static 3D MHD solution reproduce many of the characteristics derived from the fully time-dependent simulation. Despite the absence of intrinsic coronal evolution, the resulting backgrounds exhibit qualitatively similar morphology and quantitatively comparable performance in terms of both signal loss and structural preservation. In fact, our comparison with the underlying 3D coronal density showed that, although the streamer belt undergoes continuous intrinsic evolution, the large-scale projected density distribution remains sufficiently stable over multiple days that the same rarefied coronal structures continue to dominate the constructed backgrounds. Although this result has no direct analogue for real coronagraph observations, it has important implications for numerical modelling. In particular, it suggests that, for many applications, steady-state coronal models combined with realistic observing geometries may provide a practical and computationally inexpensive tool for generating high-fidelity synthetic coronagraph imagery that can be processed in the same way as real observations, thus enabling more direct and meaningful comparisons.

Finally, we emphasise that the interval analysed in this work corresponds to the near-maximum phase of the solar cycle, during which the heliospheric current sheet is highly warped and the large-scale coronal magnetic field evolves relatively rapidly. Despite these intrinsically dynamic conditions, the background subtraction methods examined here preserve the overall morphology of the K-corona remarkably well over the three-day interval considered. Whether the quantitative results reported here depend on the phase of the solar cycle remains an open question. Although the slower intrinsic evolution of the corona during solar minimum might intuitively be expected to improve the performance of background subtraction, the much flatter heliospheric current sheet means that the streamer belt remains confined to a relatively narrow range of PAs over a solar rotation. Consequently, the balance between dense streamer regions and projected open-field regions contributing to the constructed backgrounds may differ substantially from the solar maximum conditions investigated here. A similar analysis during solar minimum would therefore provide valuable insight into the extent to which the conclusions presented here depend on the global coronal magnetic configuration and its time-dependent evolution.

Overall, our results demonstrate that the large-scale morphology of the K-corona remains largely intact following background subtraction, although its photometric properties are inevitably modified to varying degrees. Quantifying these effects is therefore essential for studies that rely on the absolute or relative brightness of coronal structures. \editone{An important limitation of the present study is that the synthetic observations contain only the K-corona and do not account for the effects of the overlying F-corona on background construction and subtraction. While this idealised setup allowed us to isolate the impact of background removal on the K-coronal signal, the results do not capture the full complexity of processing real coronagraph data. Building on the foundation laid by this work, our next step will be to incorporate a modelled F-corona into the MHD-generated synthetic imagery, allowing us to investigate how the combined K- and F-coronal signals affect background construction and subtraction, and how these effects may differ from those found for the ``dust-free'' corona analysed here. We will then introduce additional sources of image degradation, including random noise, missing pixels, and instrumental artifacts. This stepwise approach will allow us to test the background subtraction procedure under progressively more realistic observing conditions, while quantifying simultaneously how effectively the F-corona is removed and how much K-coronal signal is lost in the process.}



\section*{Acknowledgements}

E.P., C.D., R.D., R.M.C., and J.A.L.\ acknowledge support from NSF's SHINE (grant no.\ AGS-2301403) programme as well as NASA's HSR (grant no.\ 80NSSC25M7101) and LWS-SC (grant no.\ 80NSSC22K0893) programmes.
P.H.\ is supported by the Office of Naval Research and NASA.
B.J.L.\ acknowledges support from NASA XRP (grant no.\ 80NSSC25K7545), LWS (grant no.\ 80NSSC24K1121), and GSFC (grant no.\ 80NSSC26K0467).
The MHD simulations analysed in this work were performed using NASA's Advanced Supercomputing Division (NAS) allocations on the Pleiades, Electra, and Aitken supercomputers, as well as NSF's ACCESS allocations on the Expanse supercomputer at the San Diego Supercomputer Center (SDSC).

\bibliography{bibliography}{}
\bibliographystyle{aa}


\appendix
\nolinenumbers

\section{The sequences of daily backgrounds} \label{app:backgrounds}

\begin{figure*}[p!]
\centering
\includegraphics[width=.83\linewidth]{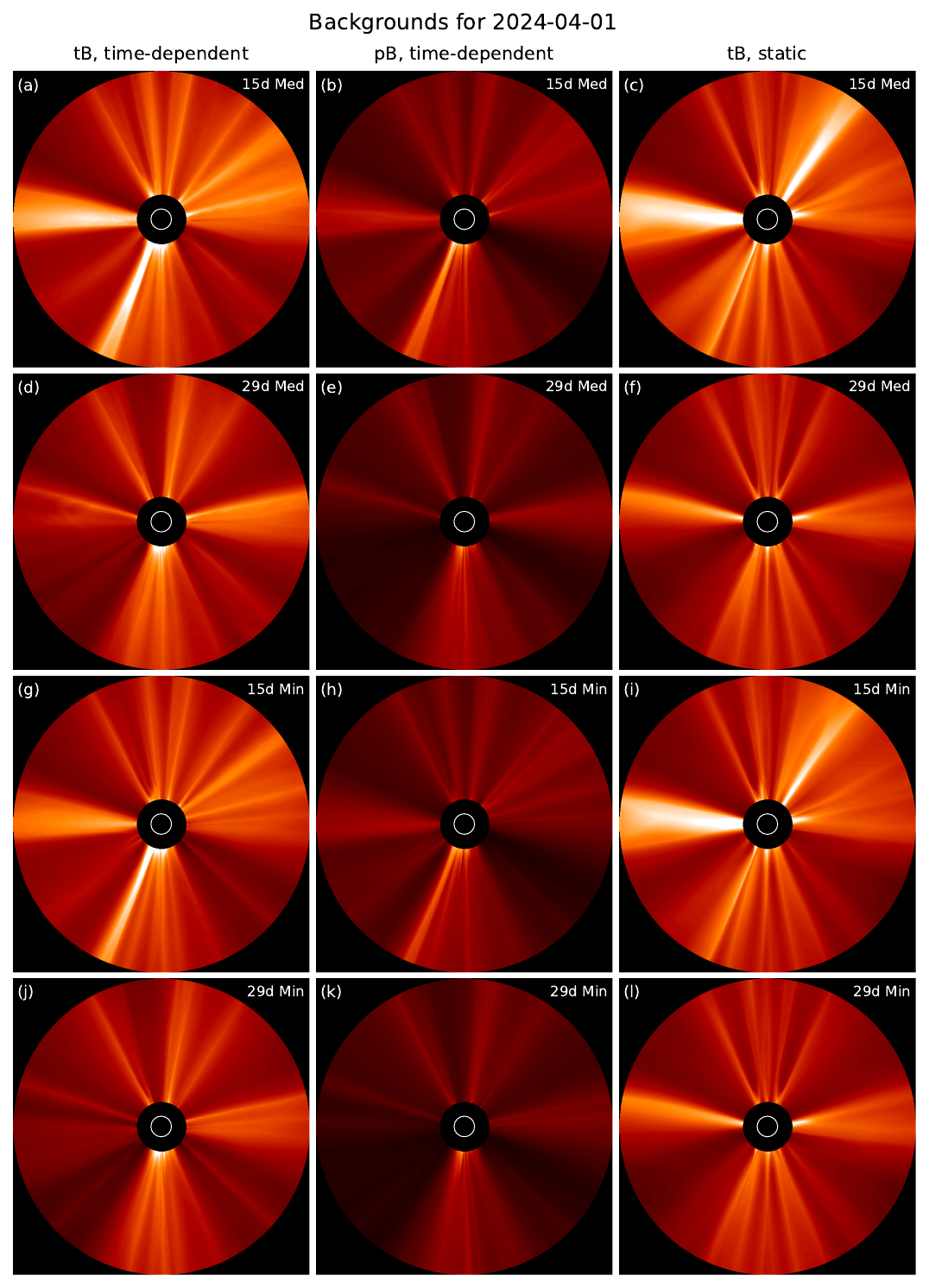}
\caption{All the daily backgrounds generated and employed for April 1, 2024. The columns display backgrounds based on (left) time-dependent tB, (middle) time-dependent pB, and (right) static tB raw imagery. To better highlight intrinsic differences in the average values across the different data sets, each panel is normalised to the same intensity data range. The animated version of the figure shows how the different daily backgrounds change in time over the three days considered in this work (from March 31 to April 2, 2024).}
\label{fig:backgrounds_20240401}
\end{figure*}

As mentioned in Section~\ref{sec:setup}, in this work we employ four background construction methods: the minimum of 29 daily medians (29d Med), the minimum of 29 days of data (29d Min), the minimum of 15 daily medians (15d Med), and the minimum of 15 days of data (15d Min). Note that each of these series is centred on the day of interest, e.g., the 29-day backgrounds for April 1, 2024 include data between March 18 and April 16, whilst the corresponding 15-day backgrounds span the interval from March 25 to April 9. Given that the whole simulation covers slightly over 32 days between March 16 and April 17, 2024, we are able to generate backgrounds for three total days: March 31, April 1, and April 2. The daily backgrounds for April 1, 2024, for the three data sets (time-dependent tB and pB as well as static tB) are displayed in Figure~\ref{fig:backgrounds_20240401}, and the full sequence of imagery including the remaining two days is provided in the animated version of the figure. We remark that these data are displayed with identical intensity ranges to highlight intrinsic differences in coronal emission. Because polarised scattering efficiency is more sensitive to the location of the source and drops off more sharply as the scattering deviates from the Thomson surface, the tB signal will include photons scattered from a wider array of locations along a given line of sight. Therefore, the pB intensity is always less than the tB intensity, which is why the pB images appear consistently darker in Figure~\ref{fig:backgrounds_20240401} \citep{deforest2013}.

Overall, it is clear from the animated version of Figure~\ref{fig:backgrounds_20240401} that the the 15-day backgrounds (regardless of construction method or raw data set) appear more variable across the interval considered than the 29-day ones. This is to be expected, since the contribution of a single image to the overall background is approximately doubled in the former case---meaning that changes over time scales of a day will have a larger impact on the final result. It is also possible to notice that the Med images tend to be slightly brighter than their corresponding Min ones, which is also to be expected due to the intrinsic background construction method (i.e., the minimum of a series of medians is consistently higher than a ``pure'' minimum). On average, the mean brightness value of a given Med image tends to be ${\sim}10$\% higher than its related Min background across all data sets. Finally, the static tB backgrounds appear to evolve less than the time-dependent ones over the three days considered, which is intuitively reasonable since changes are only due to the spacecraft's orbit around the Sun rather than an evolving, dynamic corona.

\section{Background intensity by position angle} \label{app:rings_bkg}

\begin{figure*}[t!]
\centering
\includegraphics[width=\linewidth]{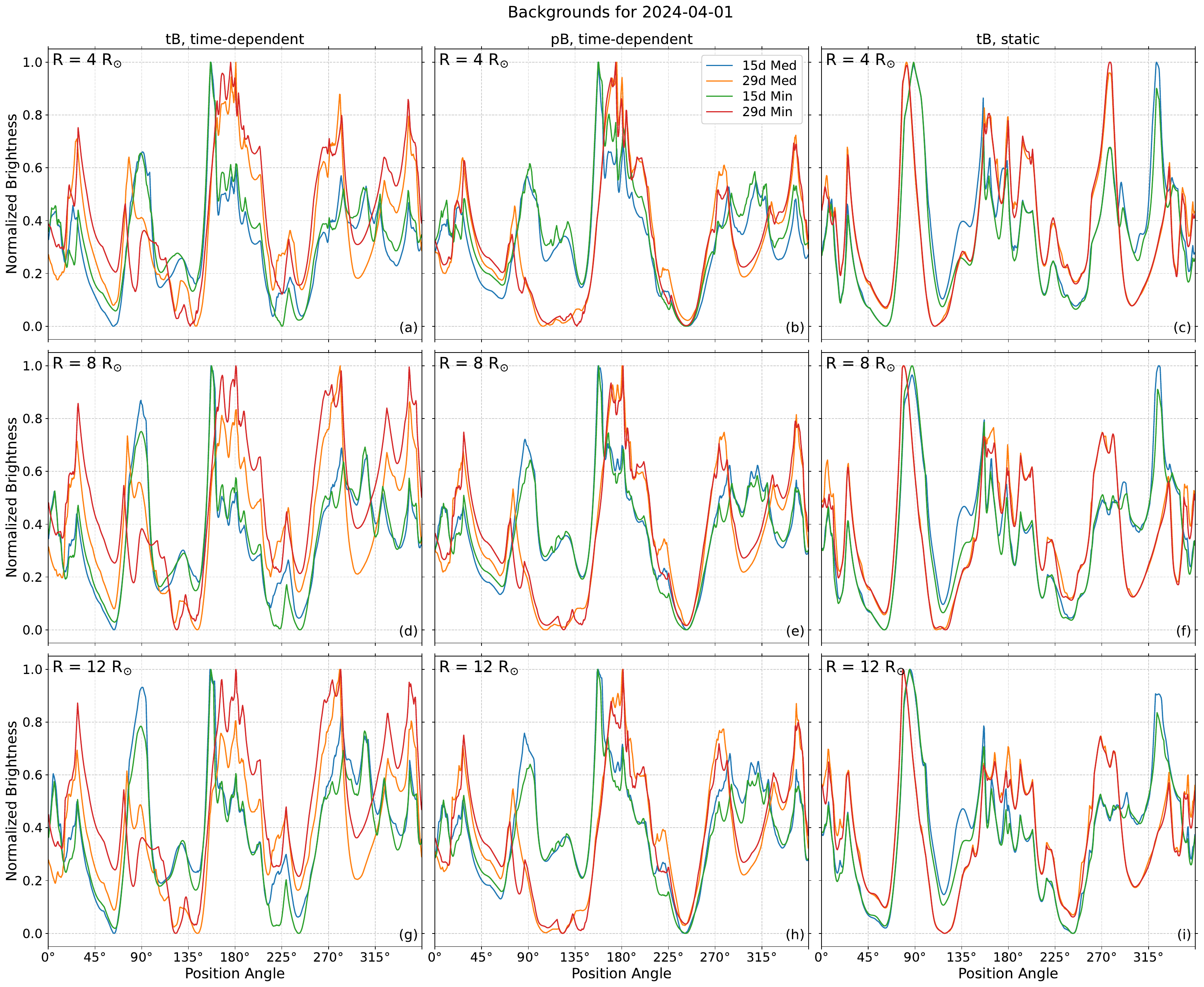}
\caption{Normalised brightness at three different coronal heights (4, 8, and 12\,$R_{\odot}$) with respect to position angle for the generated backgrounds (15d~Med, 29d~Med, 15d~Min, and 29d~Min) on April 1, 2024. The different columns show the (left) time-dependent tB, (middle) time-dependent pB, and (right) static tB data sets. The animated version of the figure shows how emission at different coronal heights for the constructed backgrounds changes in time over the three days considered in this work (from March 31 to April 2, 2024).}
\label{fig:ringplot_backgrounds}
\end{figure*}

As discussed in Section~\ref{subsec:rings}, we use annular profiles of the coronal brightness at different heliocentric heights to evaluate how closely the relative intensity of structures in background-subtracted images follows that of the corresponding raw data. A direct comparison between the background themselves in an analogous format is shown in Figure~\ref{fig:ringplot_backgrounds} for April 1, 2024 (derived from the images shown in Figure~\ref{fig:backgrounds_20240401}), and for all three days of backgrounds in the animated version of the figure. As in Figure~\ref{fig:ringplot_images}, the annular profiles are normalised to emphasise changes in the relative brightness of coronal structures rather than their absolute intensities.

Overall, the two 29-day backgrounds are qualitatively very similar to one another, as are the two 15-day counterparts. The most prominent difference arise from the length of the background window rather than from the choice of background-construction method (Med versus Min). Although the 15- and 29-day backgrounds share the same large-scale coronal morphology, they exhibit several noticeable differences in the relative brightness of individual structures. For example, the tB profiles in Figure~\ref{fig:ringplot_backgrounds} display two prominent streamer peaks between PAs of approximately $250^{\circ}$ and $360^{\circ}$ in the 29-day series, whereas the corresponding 15-day profiles are considerably flatter and feature several smaller peaks in emission. Likewise, the time-dependent pB profiles show a moderately bright structure between PAs of approximately $50^{\circ}$ and $150^{\circ}$ in the 15-day series, while the same region corresponds to one of the faintest parts of the 29-day backgrounds. More generally, the relative brightness of the major streamer structures differs systematically between the two time intervals, indicating that the length of the background window influences not only the overall brightness level but also the relative weighting of individual coronal features.

This behaviour is even more apparent in the animated version of Figure~\ref{fig:ringplot_backgrounds}, which illustrates how much more rapidly the 15-day background profiles evolve from one day to the next compared to their 29-day counterparts. This confirms the conclusions drawn from visual inspection of the animated version of Figure~\ref{fig:backgrounds_20240401}, where the 15-day backgrounds exhibited noticeably larger day-to-day variations. These results indicate again that the traditional 29-day backgrounds, spanning approximately one solar rotation, provide a more temporally consistent representation of the quiescent corona. Because each daily background is derived from a longer time interval, the replacement of the oldest day by the newest has only a modest effect on the resulting background, leading to a much smoother temporal evolution than in the 15-day case.

Finally, another notable feature is the close agreement between the backgrounds derived from the time-dependent tB and pB data sets across all heliocentric distances. Their normalised PA profiles are generally much more similar to one another than to those obtained from the static tB data, whose relative brightness distribution exhibits larger departures owing to the lack of temporal evolution in the sequence of images.

\end{document}